\documentclass{article}
\usepackage{a4wide}
\usepackage[utf8]{inputenc}
\usepackage{tikz}
\usetikzlibrary{shapes, arrows, positioning, calc, fit}
\usepackage{xcolor,colortbl}
\usepackage{subfigure}
\usepackage{stfloats}
\usepackage{siunitx}
\usepackage{authblk}
\usepackage{amsmath}
\usepackage{dirtytalk}

\title{Refining Epidemiological Forecasts with Simple Scoring Rules}

\author[]{Robert E. Moore, Conor Rosato and Simon Maskell}

\affil[]{Department of Electrical Engineering and Electronics, University of Liverpool, Brownlow Hill, Liverpool, L69 3GJ, UK}

\date{March 2022}

\begin{document}

\maketitle

\begin{abstract}
Estimates from infectious disease models have constituted a significant part of the scientific evidence used to inform the response to the COVID-19 pandemic in the UK. These estimates can vary strikingly in their bias and variability. Epidemiological forecasts should be consistent with the observations that eventually materialise. We use simple scoring rules to refine the forecasts of a novel statistical model for multisource COVID-19 surveillance data by tuning its smoothness hyperparameter.
\end{abstract}

\section{Introduction}

Several epidemiological modelling groups use statistical models of infectious disease to generate forecasts that contribute to a body of scientific evidence that informs the response to the COVID-19 pandemic in the UK. The models developed by the University of Cambridge MRC Biostatistics Unit and Public Health England (PHE) \cite{1}, the University of Warwick \cite{2}, and the London School of Hygiene \& Tropical Medicine (LSHTM) \cite{3} provide three notable examples of statistical models used to produce such estimates.

Although Cramer et al.~\cite{4} and Funk et al.~\cite{5} consider the assessment of quantile-format forecasts for COVID-19, it does not appear to be standard practice to assess full-distribution epidemiological forecasts by comparing them to the observations that eventually materialise. Maishman et al.~\cite{6} provide a set of anonymised estimates for the effective reproduction number $R_t$ that highlights the striking differences in the bias and variability of estimates that different epidemiological models can produce. We contribute to the collective effort of modelling COVID-19 in this paper by introducing a statistical model for multisource COVID-19 surveillance data and by using simple scoring rules to refine its forecasts and improve its predictive performance. The statistical model is novel by its use of symptom report data from the NHS 111 telephone service, its compartments for convalescing and terminally ill individuals, and its implementation, which uses a bespoke numerical integrator to solve the system of ODEs for the transmission model.

We begin by describing the novel statistical model in Section~\ref{sec:statistical_model} and then proceed to define a set of simple scoring rules in Section~\ref{sec:scoring_rules}. In Section~\ref{sec:computational_experiments}, we use the simple scoring rules to refine the forecasts of the statistical model. Finally, in Section~\ref{sec:conclusion}, we bring the paper to an end by presenting our conclusions.

\section{Statistical Model}\label{sec:statistical_model}

The model we describe in this section builds on a previous version whose implementation we contributed to the CoDatMo (Covid Data Models) organisation on GitHub \cite{7}. The purpose of CoDatMo is to provide a collection of COVID-19 models, all written in the statistical modelling language Stan \cite{8}. In addition to models originally implemented in Stan, CoDatMo provides Stan instantiations of COVID-19 models, initially written in other programming languages. By hosting a set of well-documented models, all implemented in a common language, CoDatMo hopes to improve understanding of the set of existing models and make it easier for newcomers and established researchers within the domain of epidemiology to make extensions and potential improvements.

We note that CoDatMo has already had some success against its objectives, with a group, primarily based at Universidade Nove de Julho in Brazil, creating a related but simpler model \cite{9}. We also note that the UK Health Security Agency uses a slightly more sophisticated version of the model presented in this paper to generate weekly estimates of the effective reproduction number and growth rate for regions of the UK \cite{10}.

The statistical model consists of two parts: a transmission model that captures a simplified mechanism for the spread of coronavirus through the population; and an observation model that encapsulates the assumptions about the connection between the states of the transmission model and the observed surveillance data used to calibrate the model.

\subsection{Transmission Model}\label{sec:transmission_model}

The simple SIR compartmental model developed by Kermack and McKendrick \cite{11} provides the theoretical basis for the transmission model. As with the SIR model, we assume a single geographical region with a large population of identical individuals who come into contact with one another uniformly at random but do not come into contact with individuals from other areas. In contrast to some other models, for example, those developed by Birrell et al. \cite{1} and Keeling et al. \cite{2}, we treat the population as identical in terms of age and sex and only discriminate between individuals based on their disease states. We also assume that the population is closed, meaning that no births or deaths occur and no migration in or out of the population occurs.

The SIR model divides the population into three disease states: individuals who are (S) susceptible to infection; individuals who have been infected with the disease and are (I) infectious to other people; and individuals who have (R) \say{recovered} either by recuperating from the disease or dying. We augment these compartments in the transmission model by adding disease states for individuals who have been (E) exposed to the virus but are not yet infectious, individuals that are no longer infectious and whose final disease states are (P) pending, and individuals who have (D) died of the disease. The exposed and dead compartments are standard extensions to the original SIR model. In contrast, we believe the pending compartments to be a novel aspect of the model. These compartments contain individuals who are either convalescing or terminally ill as a result of infection. In addition to adding these compartments, we redefine the (R) recovered population to include the living only.

Inspired by the work of the University of Cambridge MRC Biostatistics Unit and Public Health England (PHE) \cite{1}, we partition each of the intermediate disease states (E, I, P) into two sub-states. Partitioning the sub-states in this way makes the model more realistic by implicitly constraining the times spent in each of these disease states to have Erlang rather than Exponential distributions.

We assume that there is at least one individual in each susceptible, exposed, and infectious compartment on the 17th of February 2020, which is the beginning of time in the model. At this stage, we assume that no population members are pending, have recovered, or have died. Two parameters, $\alpha_1$ and $\alpha_2$, determine the allocation of the rest of the population to the first five compartments of the transmission model. $\alpha_1$  is the proportion of the remaining population initially in the susceptible compartment, and $\alpha_2$  is the proportion of the infected population that is not yet infectious at time zero. For simplicity, we divide the exposed and infectious populations equally between the respective sub-states. We can express the exact relationship between the two parameters, $\alpha_1$ and $\alpha_2$, and the initial state of the transmission model as a set of equations:

\begin{align}\label{eq:initial_state}
   S\left(0\right) &= \left(N-5\right)\alpha_1 + 1, \\
   E_1\left(0\right) &= \frac{1}{2}\left(N-5\right)\left(1-\alpha_1\right)\alpha_2 + 1, \\
   E_2\left(0\right) &= \frac{1}{2}\left(N-5\right)\left(1-\alpha_1\right)\alpha_2 + 1, \\
   I_1\left(0\right) &= \frac{1}{2}\left(N-5\right)\left(1-\alpha_1\right)\left(1-\alpha_2\right) + 1, \\
   I_2\left(0\right) &= \frac{1}{2}\left(N-5\right)\left(1-\alpha_1\right)\left(1-\alpha_2\right) + 1, \\
   P_1\left(0\right) &= 0, \\
   P_2\left(0\right) &= 0, \\
   R\left(0\right) &= 0, \\
   D\left(0\right) &= 0.
\end{align}

\begin{figure}[!h]
\centering
\includegraphics[width=\linewidth]{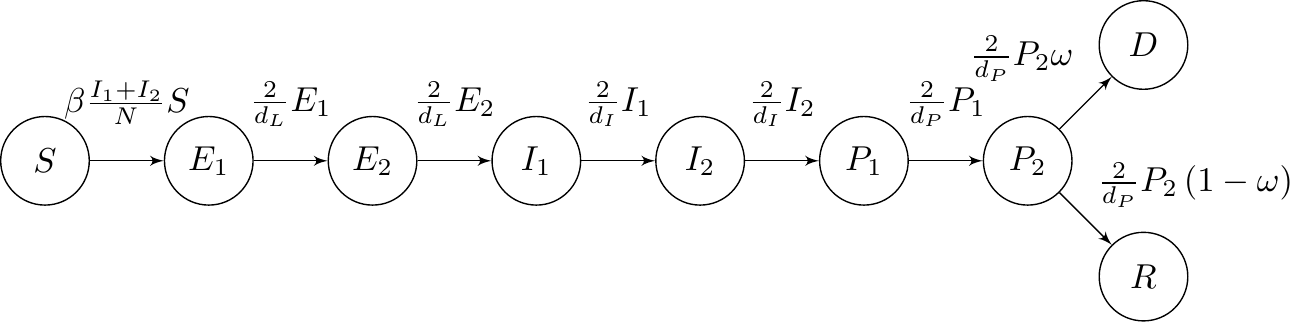}
\caption{A graph of the transmission model. Individuals begin their journey in the susceptible (S) state. From here, they are infected and move into the exposed (E) state. After the virus has incubated for a while, they continue into the infectious (I) state. Next, they enter the pending (P) state, after which they either migrate into the recovered (R) state if convalescing or pass into the deceased (D) state if terminally ill.}
\label{fig:transmission_model}
\end{figure}

Figure~\ref{fig:transmission_model} provides a graphical illustration of the transmission model that captures the assumptions relating to the flow of individuals between disease states.

We assume that the population randomly mixes as time elapses, with infectious and susceptible individuals coming into contact with one another, potentially transmitting the virus. Susceptible people who have become exposed through these contacts are not initially infectious. The virus replicates in their bodies for a time, known as the latent period, before they become infectious and have the potential to transmit the virus onto members of the remaining susceptible population. After being infectious for some time, we assume that individuals enter a state of pending before either recovering and becoming indefinitely immune to reinfection if they were convalescing or dying if they were terminally ill.

The number of individuals in each disease state varies with time according to a system of ordinary differential equations:

\begin{align}\label{eq:system_of_odes}
    \frac{dS(t)}{dt} &= -\beta(t) \frac{I_1(t) + I_2(t)}{N} S(t), \\
    \frac{dE_1(t)}{dt} &= \beta(t) \frac{I_1(t) + I_2(t)}{N} S(t) - \frac{2}{d_L} E_1(t), \\
    \frac{dE_2(t)}{dt} &= \frac{2}{d_L} \big[E_1(t) - E_2(t)\big], \\
    \frac{dI_1(t)}{dt} &= \frac{2}{d_L} E_2(t) - \frac{2}{d_I} I_1(t), \\
    \frac{dI_2(t)}{dt} &= \frac{2}{d_I} \big[I_1(t) - I_2(t)\big], \\
    \frac{dP_1(t)}{dt} &= \frac{2}{d_I} I_2(t) - \frac{2}{d_P} P_1(t), \\
    \frac{dP_2(t)}{dt} &= \frac{2}{d_P} \big[P_1(t) - P_2(t)\big], \\
    \frac{dR(t)}{dt} &= \frac{2}{d_P} P_2(t) \big[1 - \omega\big], \\
    \frac{dD(t)}{dt} &= \frac{2}{d_P} P_2(t) \omega,
\end{align}

\noindent where

\begin{itemize}
    \item $S(t)$ is the number of susceptible individuals who have not yet been infected and are at risk of infection,
    \item $E_1(t) + E_2(t)$ is the number of exposed individuals who have been infected but are not yet infectious,
    \item $I_1(t) + I_2(t)$ is the number of infectious individuals,
    \item $P_1(t) + P_2(t)$ is the number of pending individuals who are either convalescing or are terminally ill,
    \item $R(t)$ is the number of recovered individuals,
    \item $D(t)$ is the number of dead individuals,
    \item $N = S(t) + E_1(t) + E_2(t) + I_1(t) + I_2(t) + P_1(t) + P_2(t) + R(t) + D(t)$ is the constant total number of individuals in the population,
    \item $d_L$ is the mean time between infection and onset of infectiousness,
    \item $d_I$ is the mean time for which individuals are infectious,
    \item $d_P$ is the mean time for which individuals are pending,
    \item $\omega$, the infection fatality ratio (IFR), is the proportion of infected individuals who will die,
    \item $\beta(t)$ is the mean rate of contacts between individuals per unit time that are sufficient to lead to transmission if one of the individuals is infectious and the other is susceptible. $\beta(t)$ is a continuous piecewise linear function of time:
\end{itemize}

\begin{equation}
    \beta(t) = \sum_{j=1}^{J} \beta_j(t) \chi_{[t_{j-1}, t_j)}(t),
\end{equation}

\noindent where the mean rate of effective contacts during the $j$th time interval, $\beta_j(t)$, is given by

\begin{equation}\label{eq:effective_contact_rate_interval_j}
    \beta_j(t) = \frac{\beta_{j+1} - \beta_j}{t_j - t_{j-1}}(t - t_{j-1}) + \beta_j,
\end{equation}

\noindent and

\begin{equation}\label{eq:indicator_function}
    \chi_{[t_{i-1}, t_i)}(t) = \begin{cases}
    1 & \mathrm{if} \, t \in [t_{i-1}, t_i), \\
    0 & \mathrm{if} \, t \notin [t_{i-1}, t_i).
    \end{cases}
\end{equation}

The effective contact rate parameters $\beta_1, \beta_2, ..., \beta_{J+1}$ in Equation~\ref{eq:effective_contact_rate_interval_j} are the values $\beta(t)$ takes on a set of predefined dates $t_0, t_1, ..., t_J$. The first date is the 17th of February 2020, and each date that follows is seven days after the last, with the second date being the 24th of March 2020, the first day after the Prime Minister announced the first national lockdown.

\subsection{Observation Model}\label{sec:observation_model}

Epidemiological modelling groups use many types of surveillance data to calibrate statistical models of infectious diseases. The observation model captures the assumptions about the relationship between the states of the transmission model and the surveillance data that we use for calibration.

We have designed the observation model to be extensible. Here, the model only has components for death, hospital admission and symptom report data to keep things simple. Nonetheless, the observation model can be extended to assimilate additional types of surveillance data, such as case data, by appending extra components similar in structure to those for ingesting the hospital admission and symptom report data.

\subsubsection{Death Data}\label{sec:death_data}

On their official website for coronavirus data \cite{12}, the UK government publishes a daily time series of the number of deaths of individuals whose death certificate mentioned COVID-19 as one of the causes. We assume that the number of deaths on day $t$, according to this definition, $d_{\mathrm{obs}}\left(t\right)$, has a negative binomial distribution parameterised by a mean $d\left(t\right)$ and parameter $\phi_{\mathrm{deaths}}$ which affects overdispersion:

\begin{equation}\label{eq:observed_deaths}
d_{\mathrm{obs}}\left(t\right) \sim \text{NegativeBinomial}\left(d\left(t\right), \phi_{\mathrm{deaths}}\right),   
\end{equation}

\noindent where we use the alternative parameterisation of the negative binomial distribution as defined by the Stan Development Team \cite{13}:

\begin{equation}\label{eq:negative_binomial}
\text{NegativeBinomial}(\mu, \phi)=\left(\begin{array}{c}
n+\phi-1 \\
n
\end{array}\right)\left(\frac{\mu}{\mu+\phi}\right)^{n}\left(\frac{\phi}{\mu+\phi}\right)^{\phi}.
\end{equation}

In Equation~\ref{eq:observed_deaths}, $d\left(t\right)$ is the difference between the population of the D state of the transmission model between days $t-1$ and $t$: $d\left(t\right) = D\left(t\right) - D\left(t-1\right)$.

\subsubsection{Hospital Admission Data}\label{sec:hospital_admission_data}

The UK government also publishes a daily time series of the number of COVID-19 patients admitted to hospital on their official website for coronavirus data~\cite{12}. We assume that, like the number of deaths, the number of hospital admissions on day $t$, $h_{\text{obs}}(t)$, has a negative binomial distribution parameterised by $h(t)$ and $\phi_{\text{admissions}}$:

\begin{equation}\label{eq:observed_hospital_admissions}
    h_{\mathrm{obs}}\left(t\right) \sim \text{NegativeBinomial}\left(h\left(t\right), \phi_{\mathrm{admissions}}\right),   
\end{equation}

\noindent where

\begin{equation}
    h\left(t\right) = \rho_\text{admissions}\left(t\right) \times \frac{2}{d_I} I_2,
\end{equation}

\noindent i.e. the mean number of hospital admissions on day $t$ equals the ratio of hospital admissions to potential patients, $\rho_{\text{admissions}}(t)$, multiplied by the number of new members of the pending state. $\rho_{\text{admissions}}(t)$ is a continuous piecewise linear function of time:

\begin{eqnarray}\label{eq:ratio_of_hospital_admissions_to_potential_patients}
    \rho_{\text {admissions}}(t) = \sum_{k=1}^{K} \rho_{\text{admissions}, k}(t) \chi_{\left[t_{k-1}, t_{k}\right)}(t),
\end{eqnarray}

\noindent where the ratio of hospital admissions to potential patients during the $k$th time interval, $\rho_{\text{admissions}, k}(t)$, is given by

\begin{equation}\label{eq:ratio_of_hospital_admissions_to_potential_patients_interval_k}
    \rho_{\text{admissions}, k}(t) = \frac{\rho_{\text{admissions}, k+1} - \rho_{\text{admissions}, k}}{t_k - t_{k-1}}(t - t_{k-1}) + \rho_{\text{admissions}, k},
\end{equation}

\noindent and the indicator function for the $k$th time interval, $\chi_{\left[t_{k-1}, t_{k}\right)}(t)$, is defined in Equation~\ref{eq:indicator_function}.

The parameters $\rho_{\text{admissions}, 1}$, $\rho_{\text{admissions}, 2}$, ..., $\rho_{\text{admissions}, K+1}$ in Equation~\ref{eq:ratio_of_hospital_admissions_to_potential_patients_interval_k} are the values $\rho_{\text {admissions}}(t)$ takes on a set of predefined dates $t_0$, $t_1$, ..., $t_K$. The first date is the 24th of March 2020, and each date that follows is twelve weeks after the last, with the second date being the 16th of June 2020.

\subsubsection{Symptom Report Data}\label{sec:symptom_report_data}

Every weekday up to the previous calendar day, NHS Digital publishes a daily time series of the number of assessments completed through the NHS 111 telephone service where callers reported potential coronavirus (COVID-19) symptoms \cite{14}. Leclerc et al. found a strong correlation between the volume of these symptom reports and the number of COVID-19 deaths reported 16 days later \cite{15}. We assume that, like the other types of surveillance data, the number of assessment calls to NHS 111 on day $t$ where callers reported potential COVID-19 symptoms, $c_{\mathrm{obs}}\left(t\right)$, has a negative binomial distribution parameterised by $c\left(t\right)$ and $\phi_{\mathrm{calls}}$:

\begin{equation}
    c_{\mathrm{obs}}\left(t\right) \sim \text{NegativeBinomial}\left(c\left(t\right), \phi_{\mathrm{calls}}\right),
\end{equation}

\noindent where

\begin{equation}
    c\left(t\right) = \rho_\text{calls}\left(t\right) \times \left(\frac{2}{d_L} E_2 + \frac{2}{d_I} I_2\right),
\end{equation}

\noindent i.e. the mean number of assessment calls to NHS 111 on day $t$ where callers reported potential COVID-19 symptoms equals the ratio of symptom reports to potential symptom reporters, $\rho_{\text{calls}}(t)$, multiplied by the sum of the number of new members of the infectious and pending states.

$\rho_{\text{calls}}(t)$ is a continuous piecewise linear function of time almost identical to $\rho_{\text{admissions}}(t)$, which is defined by Equations~\ref{eq:ratio_of_hospital_admissions_to_potential_patients} and~\ref{eq:ratio_of_hospital_admissions_to_potential_patients_interval_k}. The only difference is that the parameters $\rho_{\text{calls}, 1}$, $\rho_{\text{calls}, 2}$, ..., $\rho_{\text{calls}, L+1}$ are associated with a different set of predefined dates $t_0$, $t_1$, ..., $t_L$. The first of these dates is the 24th of March 2020, and each date that follows is four weeks after the last, with the second date being the 16th of March 2020.

\section{Scoring Rules}\label{sec:scoring_rules}

Scoring rules produce real numbers, also called numerical scores, that summarise the quality of probabilistic forecasts. More concretely, consider a probabilistic forecast $P$ of an uncertain future quantity $X$ for which the observation $x$ eventually materialises. In a scenario such as this, a scoring rule provides a numerical score $s(P, x)$ that quantifies the statistical consistency between the predictive distribution $P$ and the observation $x$. Table~\ref{tab:scoring_rules} shows the simple scoring rules that we use in this paper.

\renewcommand{\arraystretch}{1.9}
\begin{table}
\caption{The set of simple scoring rules that feature in this paper. In the definitions, $p_x$ is the probability mass of the predictive distribution for an observed count $x$, $\|p\|^{2}=\sum_{k=0}^{\infty} p_{k}^{2}$, $P_k$ is the value of cumulative predictive distribution for a count $k$, $\mathbf{1}(.)$ is the indicator function, and $\mu_P$ and $\sigma_P^2$ are the mean and variance of the predictive distribution.}
\begin{tabular}{p{0.3\linewidth} p{0.4\linewidth} p{0.2\linewidth}}
\hline
\textbf{Scoring Rule} & \textbf{Definition} & \textbf{Reference} \\ \hline
Logarithmic score & $\operatorname{logs}(P, x)=-\log p_{x}$ & Good~\cite{16} \\ \hline
Quadratic score & $\operatorname{qs}(P, x)=-2 p_{x}+\|p\|^{2}$ & Wecker~\cite{17} \\ \hline
Spherical score & $\operatorname{sphs}(P, x)=-\frac{p_{x}}{\|p\|}$ & Czado, Gneiting and Held~\cite{18} \\ \hline
Ranked probability score & $\operatorname{rps}(P, x)=\sum_{k=0}^{\infty}\left\{P_{k}-\mathbf{1}(x \leq k)\right\}^{2}$ & Epstein~\cite{19} \\ \hline
Dawid-Sebastiani score & $\operatorname{dss}(P, x)=\left(\frac{x-\mu_{P}}{\sigma_{P}}\right)^{2}+2 \log \sigma_{P}$ & Gneiting and Raftery~\cite{20} \\ \hline
Squared error score & $\operatorname{ses}(P, x)=\left(x-\mu_{P}\right)^{2}$ & Czado, Gneiting and Held~\cite{18} \\ \hline
Normalised~squared~error score & $\operatorname{nses}(P, x)=\left(\frac{x-\mu_{P}}{\sigma_{P}}\right)^{2}$ & Carroll and Cressie~\cite{21} \\ \hline
\end{tabular}
\label{tab:scoring_rules}
\end{table}

The logarithmic, quadratic, spherical, ranked probability, Dawid-Sebastiani, and squared error scores in Table~\ref{tab:scoring_rules} are negatively oriented, with better forecasts typically resulting in lower scores. These scores are also said to be proper in the sense that a forecaster minimises them when quoting their true belief. Proper scoring rules are considered essential for incentivising honest forecasting, a position argued by Gneiting and Raftery~\cite{20}.

In contrast, the normalised squared error score is improper, a quality that has resulted in it being discredited, for example, by Czado, Gneiting and Held~\cite{18}, as a tool for evaluating probabilistic forecasts. We argue that viewing it through the lens of propriety leads to an underappreciation of its unique properties, particularly its ability to distinguish between over-confidence and over-caution. Indeed, we see the normalised squared error score as a valuable diagnostic tool with advantages over proper scoring rules in certain situations.

Interestingly, the normalised squared error score is popular in the tracking and data fusion community, which has studied performance measures for evaluating estimation algorithms, such as Li and Zhao's absolute~\cite{22} and relative~\cite{23} error measures. Blasch, Rice, and Yang~\cite{24}, and Chen et al.~\cite{25} describe a now-popular relative error measure called the Normalised Estimation Error Squared (NEES), which is a generalisation of the normalised squared error score to multiple dimensions. Researchers in the community have used the NEES extensively as an easily understood approach to arguing the merits of, for example, different extensions of the Kalman filter to specific non-linear settings where the Extended Kalman Filter is routinely over-confident~\cite{26}.

Scores are typically reported as averages over multiple probabilistic forecasts, each for a distinct point in time. Following Czado, Gneiting, and Held~\cite{18}, we use uppercase to denote the mean score over several forecasts. The tables in this paper use the mean scores LogS, QS, SphS, RPS, DSS, SES and NSES.

\section{Computational Experiments}\label{sec:computational_experiments}

\subsection{Setup}

We implement the statistical model described in Section~\ref{sec:statistical_model} in Stan \cite{8}. Stan is a probabilistic programming language that allows users to articulate statistical models and calibrate them with data using a Markov chain Monte Carlo (MCMC) method called the No-U-Turn Sampler (NUTS), proposed by Hoffman and Gelman \cite{27}. Stan also provides diagnostic information, such as warnings about divergent transitions, to help users check it has sampled the posterior faithfully.

In addition to encoding the statistical model, the Stan implementation includes prior distributions for the model parameters. Table~\ref{tab:prior_distributions} gives details of the prior distributions we have chosen to use in the computational experiments, along with the reasoning behind their selection. Four of the prior distributions, for the parameters $d_L$, $d_I$, $d_P$ and $\omega$, are based on estimates from the published literature on COVID-19. We base the prior distributions for $d_L$ and $d_I$ on estimates of the time variables rather than estimates of their mean values, for which we could not identify reliable estimates. This decision results in looser prior distributions and, in turn, wider posterior distributions for these two parameters.

The software implementation, which is publically available on GitHub\footnote{https://github.com/codatmo/UniversityOfLiverpool\_PaperSubmission}, has two idiosyncrasies worthy of discussion. First, the current implementation does not use any of the integrators provided by Stan to solve the system of ordinary differential equations (ODEs) in Equation~\ref{eq:system_of_odes}. Instead, a bespoke implementation of the explicit trapezoidal method \cite{28} solves the system of ODEs for the transmission model. Anecdotally, the trapezoidal integrator significantly reduces runtime while producing acceptable numerical errors. Second, the current implementation only uses Stan's default initialisation strategy for $1/\phi_\text{deaths}$, $1/\phi_\text{admissions}$, $1/\phi_\text{calls}$, $\rho_{\text{admissions}, 1}$, ..., $\rho_{\text{admissions}, K+1}$, $\rho_{\text{calls}, 1}$, ..., and $\rho_{\text{calls}, L+1}$ by drawing values uniformly between -2  and 2 on the unconstrained parameter space. Rather than doing this for $\alpha_1$, $\alpha_2$, $\beta_1$, ..., $\beta_J$, $d_L$, $d_I$, $d_P$, and $\omega$, the implementation draws uniformly from custom intervals to prevent initialisation failures caused by unrealistic parameter values.

We calibrate the model with data for England by using NUTS to draw six independent Markov chains for each of seven smoothness hyperparameter, $\sigma_\beta$, values. Each chain draws 512 samples and discards the first 256 drawn during warmup. We calibrate over the period from the 24th of March 2020 to the 31st of December 2020 with the death, hospital admission, and symptom report data described in Section~\ref{sec:statistical_model}\ref{sec:observation_model}. Each calibration job produces a posterior distribution for the statistical model's parameters, shown in Table~\ref{tab:prior_distributions}. For each job, we generate two posterior predictive distributions for the daily number of deaths by simulating from the statistical model, first using the posterior samples for the parameters and second using only a point estimate for the parameters. We perceive that there are occasions where the use of point estimates can explain over-confident estimates. Specifically, we use the mean of the posterior samples as the point estimate for the parameters. The posterior predictive distributions span the 17th of February 2020 to the 21st of January 2021, with the last twenty-one elements being three-week forecasts for which we calculate the LogS, QS, SphS, RPS, DSS, SES and NSES.

\renewcommand{\arraystretch}{1.0}
\begin{table}
\caption{Prior distributions for the parameters of the statistical model with the rationale for their selection. We use the symbol $^+$ to indicate a distribution with its lower tail truncated at zero.}
\begin{tabular}{llp{6.8cm}}
\hline
\textbf{Parameter(s)} & \textbf{Prior Distribution} & \textbf{Comment} \\ \hline
$\alpha_1$ & $\text{Beta}\left(5.0, 0.5\right)$ & This reflects our belief that most of the population is initially susceptible. \\ \hline
$\alpha_2$ & $\text{Beta}\left(1.1, 1.1\right)$ & This reflects our uninformed beliefs about the initial division of the infected population into those who are infectious and those who are not. \\ \hline
$\beta_1$ & $\text{HalfNormal}\left(0.0, 0.5\right)$ & This is a generic, weakly informative prior inspired by the work of Gelman \cite{29}. \\ \hline
$\beta_2, ...,\beta_J$ & $\text{Normal}^+\left(\beta_{i-1}, \sigma_\beta\right)$ & This is random-walk prior with smoothness hyperparameter $\sigma_\beta$ that enforces correlation between contiguous $\beta_i$. \\ \hline
$d_L$ & $\text{Normal}^+(4.0, 3.0)$ & This is based on an estimate of the incubation period provided by Pellis et al. \cite{30}. \\ \hline
$d_I$ & $\text{Normal}^+(5.0, 4.0)$ & This is based on an estimate of the delay from onset of symptoms to hospitalisation provided by Pellis et al. \cite{30}. \\ \hline
$d_P$ & $\text{Normal}^+(13.0, 4.0)$ & This is based on an estimate of the mean delay from hospitalisation to death provided by Linton et al. \cite{31}. \\ \hline
$\omega$ & $\text{Beta}\left(5.7, 624.1\right)$ & This is based on an estimate provided by Ward et al. \cite{32}. \\ \hline
$\frac{1}{\phi_{\text{deaths}}}$, $\frac{1}{\phi_{\text{admissions}}}$, $\frac{1}{\phi_{\text{calls}}}$ & $\text{Exponential}\left(5.0\right)$ & This is a containment prior for the overdispersion parameter, which is discussed by Simpson \cite{33}. \\ \hline
$\rho_{\text{admissions}, k}$, $\rho_{\text{calls}, l}$ & $\text{Beta}\left(1.1, 1.1\right)$ & This reflects our uninformed beliefs about these ratio parameters. \\ \hline
\end{tabular}
\label{tab:prior_distributions}
\end{table}

\subsection{Results}\label{sec:results_section}

Mean scores for the forecasts that we generated with posterior samples for the parameters and point estimates for the parameters are presented in the top and bottom of Table~\ref{tab:mean_scores}, respectively. The columns contain results for different values of the smoothness hyperparameter $\sigma_\beta$ defined in Table~\ref{tab:prior_distributions}. Smaller values of $\sigma_\beta$ make the random-walk prior on the effective contact rate $\beta\left(t\right)$ tighter, causing it to vary more slowly and, if low enough, to underfit the data. Conversely, larger values of the smoothness hyperparameter loosen the random-walk prior and allow overfitting of the data if $\sigma_\beta$ is high enough.

\begin{table}
\centering
\caption{Mean scores for the three-week forecasts generated with different $\sigma_\beta$ values. Top: Simulating from the statistical model with a point estimate for the parameters. Bottom: Simulating from the statistical model with posterior samples for the parameters.}
\label{tab:mean_scores}
\begin{tabular}{llllllll}
\hline
\textbf{Scoring Rule} & \multicolumn{7}{c}{\textbf{$\sigma_\beta$}} \\
 & 0.0005 & 0.001 & 0.0025 & 0.005 & 0.01 & 0.025 & 0.05 \\
\hline
 & \multicolumn{7}{c}{\textbf{Point estimate}} \\ 
LogS &9.595 &7.403 &9.212 &9.711 &9.752 &9.260 &10.000\\
QS  &0.002 &-0.001 &0.002 &0.004 &0.005 &0.006 &0.990\\
SphS &-0.001 &-0.027 &-0.005 &-0.001 &0.000 &-0.004 &0.000\\
RPS &693.406 &176.730 &349.403 &202.763 &201.986 &124.211 &0.000 \\
DSS &27.478 &12.367 &29.167 & 28.042 &28.184 &24.997 &209338988\\
SES &665563 &81671 &268904 &89048 &89110 &33435 &1055387\\
NSES &16.908 &1.689 &19.666 &19.743 &19.895 &17.410 &209338994\\\hline\hline
& \multicolumn{7}{c}{\textbf{Posterior samples}} \\ 
LogS &9.624 &7.313 &9.195 &6.988 &6.949 &\bf{6.076} &6.200 \\
QS  &0.002 &0.000 &0.002 &0.000 &0.000 &-0.003 &\bf{-0.004} \\
SphS &-0.001 &-0.024 &-0.005 &-0.025 &-0.025 &-0.052 &\bf{-0.058} \\
RPS &696.828 &181.678 &393.543 &123.172 &122.706 &\bf{50.085} &74.223 \\
DSS &27.452 &12.356 &25.139 &11.797 &11.776 &\bf{9.992} &10.440 \\
SES &669959 &85439 &261226 &37660 &37607 &\bf{4690}&44607 \\
NSES &16.880 &1.581 &15.464 &2.399 &2.372 &\bf{0.340} &0.276 \\\hline
\end{tabular}
\vspace*{-4pt}
\end{table}

Generally, the mean scores in Table~\ref{tab:mean_scores} are lower, or closer to one in the case of NSES, for the forecasts generated with posterior samples for the parameters than for those generated with point estimates. This observation highlights the importance of allowing uncertainty propagation from statistical inference to forecasting. The differences between the values produced by the two forecasting methods are small for the lowest smoothness hyperparameter value of 0.0005 but increase with $\sigma_\beta$ until they become large and pronounced.

\begin{figure}
\centering
\subfigure{\includegraphics[width=0.95\linewidth]{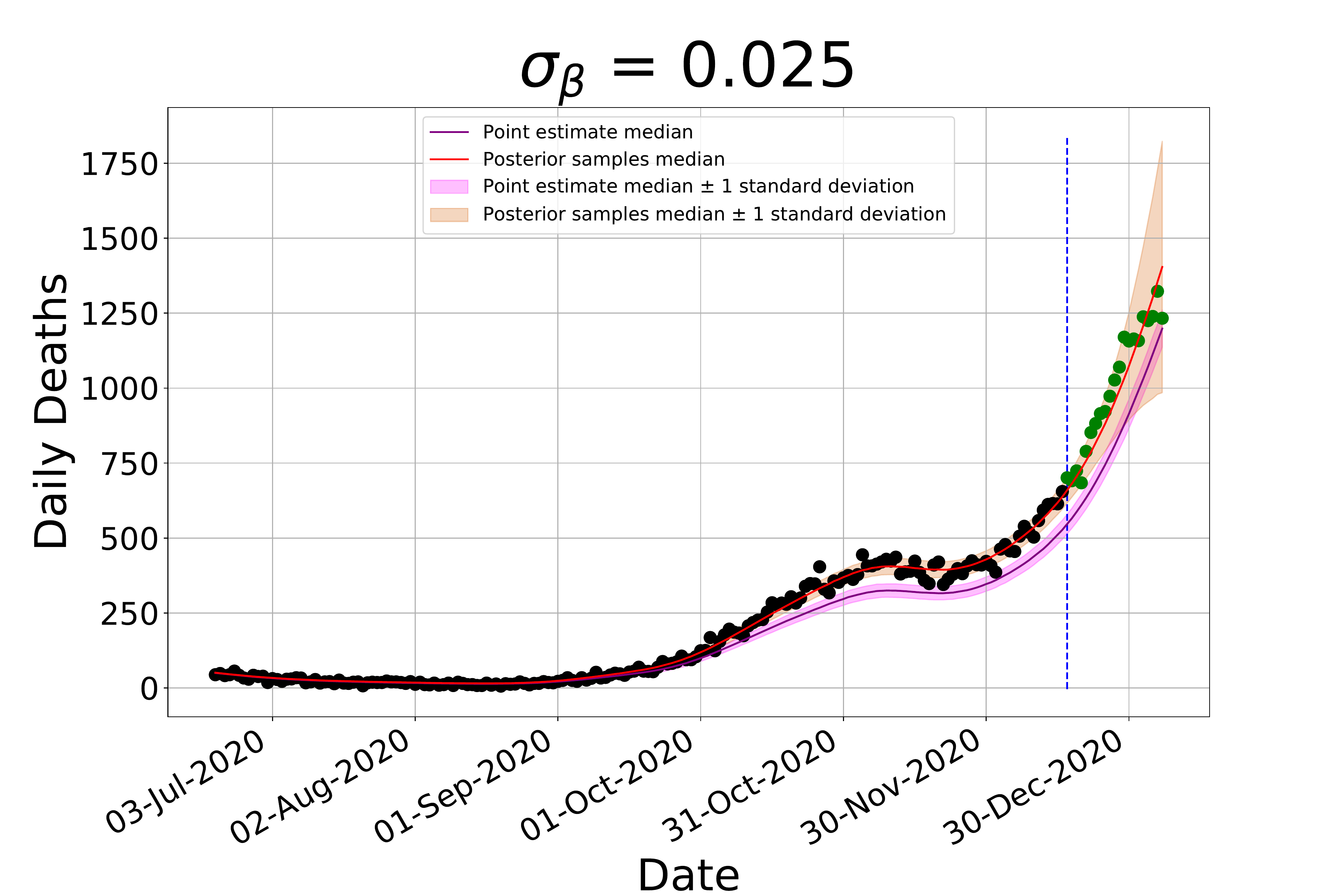}}
\hfill
\subfigure{\includegraphics[width=0.95\linewidth]{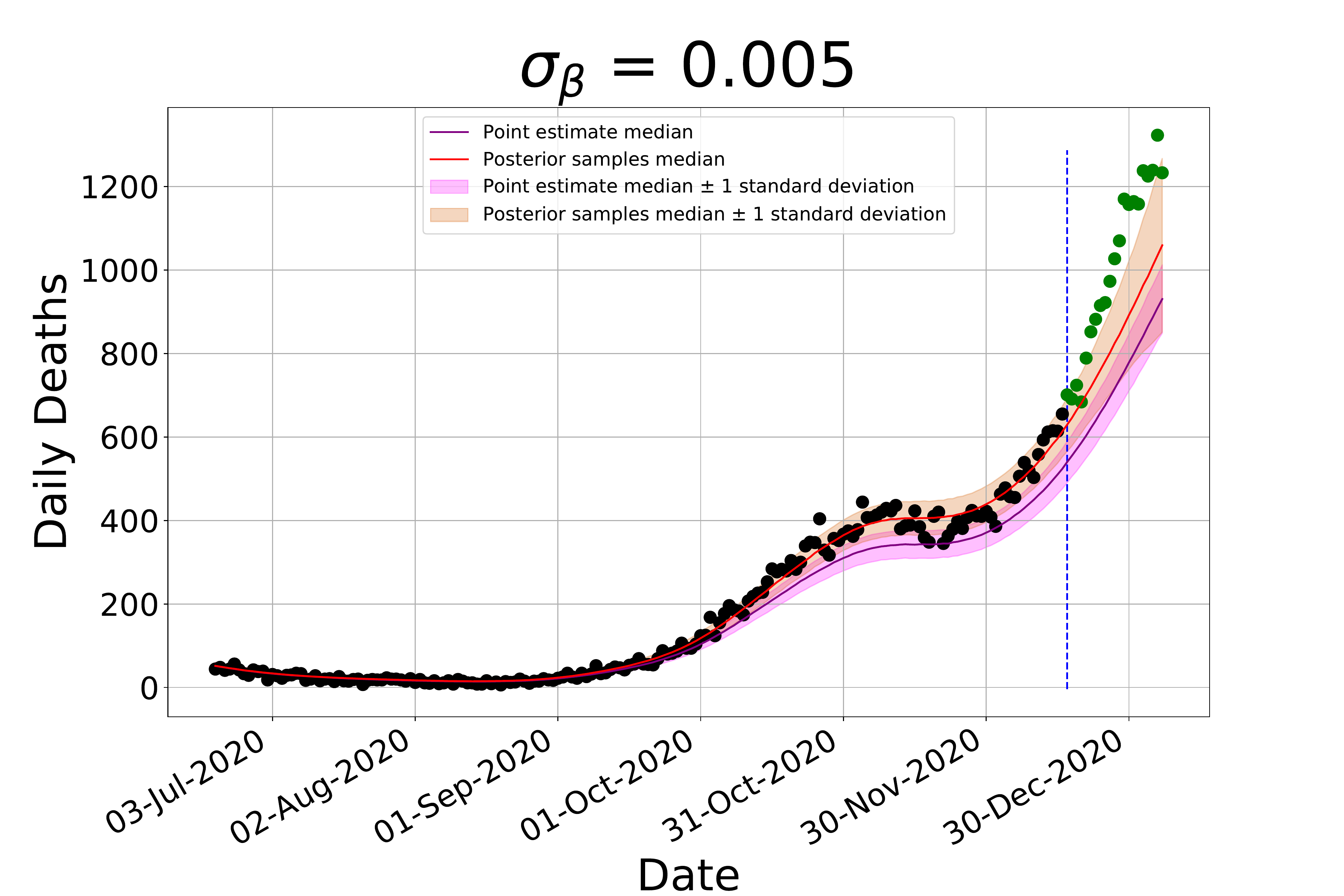}}
\caption{Forecasts generated with point estimates and posterior samples for smoothness hyperparameter, $\sigma_\beta$, values of (a) 0.025 and (b) 0.005.}
\label{fig:forecasts}
\end{figure}

We have emboldened the best value for each of the mean scores, corresponding to the lowest value for the mean proper scores, LogS, QS, SphS, RPS, DSS, and SES, and the value closest to and less than one for NSES. We select the best NSES value in this way because values of less than one correspond to over-cautious forecasts, which are arguably less damaging in terms of their impact on decision making than over-confident forecasts, which have NSES values of greater than one. The majority of the best mean scores are for the posterior samples for the parameters and a $\sigma_\beta$ value of 0.025. We can see that this forecast, shown in Figure~\ref{fig:forecasts} (a), is over-cautious, which the NESE value correctly summarises. We can also see the forecast for the point estimate for the parameters in Figure~\ref{fig:forecasts} (a). This forecast is biased because the mean resides outside the posterior distribution's region of high probability mass, as it does for the Hybrid Rosenbrock distribution~\cite{34}. Although most of the best mean scores are for a $\sigma_\beta$ value of 0.025, the best values for the QS and SphS are for a smoothness hyperparameter value of 0.05. The differences, however, between the mean scores, 0.001 for QS and 0.006 for SphS, are relatively small and indicate little difference between the quality of the two forecasts when assessed with QS and SphS.

The forecast generated with the posterior samples for the parameters and a $\sigma_\beta$ value of 0.005 has a NSES value of 2.399, which indicates that it is over-confident. We can see in Figure~\ref{fig:forecasts} (b) that the forecast fails to predict most of the future observations and is indeed over-confident. The other scoring rules do not provide any information about the over-confidence of this forecast. We, therefore, believe that NSES's ability to distinguish between over-confidence and over-caution, given a single forecast, makes it a valuable diagnostic tool that should be used alongside proper scoring rules.

\section{Conclusion}\label{sec:conclusion}

We have shown how to use simple scoring rules to develop a statistical model and improve its forecasting performance. The computational experiments presented in Section~\ref{sec:computational_experiments} demonstrate that the statistical model introduced in Section~\ref{sec:statistical_model} provides the best forecasts when we use posterior samples for the parameters and a smoothness hyperparameter $\sigma_\beta$ value of 0.025. We, therefore, advocate simple scoring rules for evaluating epidemiological forecasts and NSES specifically to establish if they are over-confident or over-cautious.

One of the significant limitations of simple scoring rules is that we can only use them to assess forecasts of observable variables. Epidemiological modellers cannot apply them to important latent quantities, such as the effective reproduction number $R_t$ and the growth rate $r$, which they often forecast. Accordingly, the epidemiological community needs a method for assessing forecasts of quantities for which the truth is unknown. Simulation-Based Calibration (SBC)~\cite{35} is a candidate method for this task, worth investigating further.

There are four worthwhile directions in which we can extend the statistical model presented in Section~\ref{sec:statistical_model}. The first direction involves adding components to the observation model described in Section~\ref{sec:observation_model} to allow calibration with a greater quantity and diversity of surveillance data. The second direction involves modifying it to accommodate surveillance data from other countries. The third direction entails making the disease-specific, geography-independent parameters $d_L$, $d_I$, $d_T$, and $\omega$ global to facilitate information sharing between regions, as is done by Birrell et al. \cite{1}. The fourth and final direction involves removing the assumption that recovered individuals are indefinitely immune to reinfection, allowing reinfection, which is more realistic.

\section*{Data Accessibility}

The code and data for the computational experiments are publically available on GitHub \\ (https://github.com/codatmo/UniversityOfLiverpool\_PaperSubmission).

\section*{Authors' Contributions}

REM designed and implemented the statistical model, participated in the analyses and drafted the manuscript; CR carried out the computational experiments, participated in the analyses and helped draft the manuscript; SM conceived and supervised the work, participated in the analyses and critically revised the manuscript. All authors gave final approval for publication and agree to be held accountable for the work performed therein.

\section*{Competing Interests}

We declare that we have no competing interests.

\section*{Funding}

This work was supported by an ICASE Research Studentship jointly funded by EPSRC and AWE [EP/R512011/1]; a Research Studentship jointly funded by EPSRC and the ESRC Centre for Doctoral Training on Quantification and Management of Risk and Uncertainty in Complex Systems Environments [EP/L015927/1]; and EPSRC through the Big Hypotheses grant [EP/R018537/1].

\section*{Acknowledgements}

We thank Public Health England (PHE), the Joint Biosecurity Centre (JBC), and the UK Health Security Agency (UKHSA) for their support. We thank Breck Baldwin and Jose Storopoli for their help in advancing CoDatMo. We thank Veronica Bowman, Alexander Phillips and John Harris for their suggestions and Matthew Carter for configuring the HPC environment. We also thank the anonymous reviewers for their insightful comments, which helped improve the paper significantly.

\end{document}